\documentclass[preprintnumbers, floatfix,letterpaper, twocolumn,aps,prd,epsfig,nofootinbib,
]{revtex4-1}
\usepackage{bm,graphicx,dcolumn,epstopdf,epsf, latexsym,mathbbol, amssymb,amsmath,color,slashed, mathrsfs,mathcomp,simplewick}
\usepackage[center]{subfigure}
\usepackage{multirow}
\usepackage{makecell}
\usepackage[colorlinks,linkcolor=blue,citecolor=blue,urlcolor=blue]{hyperref}

\begin{document}
\allowdisplaybreaks
 \newcommand{\bq}{\begin{equation}}
 \newcommand{\eq}{\end{equation}}
 \newcommand{\bqn}{\begin{eqnarray}}
 \newcommand{\eqn}{\end{eqnarray}}
 \newcommand{\nb}{\nonumber}
 \newcommand{\lb}{\label}
 \newcommand{\f}{\frac}
 \newcommand{\p}{\partial}
\newcommand{\PRL}{Phys. Rev. Lett.}
\newcommand{\PLB}{Phys. Lett. B}
\newcommand{\PRD}{Phys. Rev. D}
\newcommand{\CQG}{Class. Quantum Grav.}
\newcommand{\JCAP}{J. Cosmol. Astropart. Phys.}
\newcommand{\JHEP}{J. High. Energy. Phys.}
 \newcommand{\Doi}{https://doi.org}

\title{Primordial non-Gaussianity and power asymmetry with quantum gravitational  effects in loop quantum cosmology}

\author{Tao Zhu${}^{a, b}$}
\email{tao$\_$zhu@baylor.edu}

\author{Anzhong Wang${}^{a, b}$}
\email{anzhong$\_$wang@baylor.edu}

\author{Klaus Kirsten${}^{c}$}
\email{klaus$\_$kirsten@baylor.edu}

\author{Gerald Cleaver${}^{d}$}
\email{gerald$\_$cleaver@baylor.edu}

\author{Qin Sheng${}^{c}$}
\email{qin$\_$sheng@baylor.edu}

\affiliation{${}^{a}$ Institute for Advanced Physics $\&$ Mathematics, Zhejiang University of Technology, Hangzhou, 310032, China
\\${}^{b}$ GCAP-CASPER, Physics Department, Baylor University, Waco, TX 76798-7316, USA
\\
${}^{c}$ GCAP-CASPER, Mathematics Department, Baylor University, Waco, TX 76798-7328, USA\\
${}^{d}$ EUCOS-CASPER, Physics Department, Baylor University, Waco, TX 76798-7316, USA
}

\date{\today}

\begin{abstract}

Loop quantum cosmology (LQC) provides a resolution of the classical big bang singularity in the deep Planck era. The evolution,
prior to the usual slow-roll inflation, naturally generates excited states at the onset of the slow-roll inflation. It is expected that these
quantum gravitational effects could leave its fingerprints on the primordial perturbation spectrum and non-Gaussianity, and lead to 
some observational evidences in the cosmic microwave background (CMB). While the impact of the quantum effects on the primordial
perturbation spectrum has been already studied and constrained by current data, in this paper we continue studying such effects
on the non-Gaussianity of the primordial curvature perturbations.  In this paper, we present detailed and analytical calculations of
the non-Gaussianity and show explicitly that the corrections due to quantum effects are in the same magnitude of the slow-roll
parameters in the observable scales and thus are well within current observational constraints. Despite  this, we show that the
non-Gaussianity in the squeezed limit can be enhanced at superhorizon scales and further, these effects may yield a large
statistical anisotropy on the power spectrum through the Erickcek-Kamionkowski-Carroll  mechanism.

\end{abstract}
\maketitle

\section{Introduction}
\renewcommand{\theequation}{1.\arabic{equation}}\setcounter{equation}{0}

Primordial non-Gaussianities have attracted a great deal of attention in recent years, because the existence of their signatures in the 
CMB spectrum would be a powerful probe of the physics of inflation 
\cite{bartolo_primordial_2005, chen_observational_2007, chen_primordial_2010, komatsu_non-gaussianity_2009}. While the predictions 
regarding the nearly scale invariant and Gaussian part of the spectrum of primordial density fluctuations arising from inflation are highly 
degenerate for different models, the deviations from a perfect Gaussian distribution, the non-Gaussianity, have been proven to contain valuable information to distinguish among different models of inflation. In general, the amount of non-Gaussianities produced by the simplest single field slow-roll inflationary models is at a level proportional to the slow-roll parameters, i.e., $f_{\rm NL} \sim 10^{-2}$,  and thus unlikely to be observed by the next generation of CMB experiments and large scale surveys \cite{maldacena_non-gaussian_2003, chen_primordial_2010, wang_inflation_2014}. For this reason, a detection of non-Gaussianities in CMB would rule out the standard single-field slow-roll scenarios, leading to the study of exotic inflationary models or even theories with different dynamics for the generation of primordial perturbations. In addition, different patterns of non-Gaussianities which are predicted by inflationary models or background dynamics beyond the standard single-field slow-roll inflation also provide close connections to the quantum theory of gravity near the Planck scale; see \cite{chen_primordial_2010, wang_inflation_2014} for detailed reviews.

In standard calculations of non-Gaussianities, the primordial perturbation modes are in general assumed to be at the Bunch-Davies (BD) vacuum state when they were inside the Hubble horizon at the beginning of inflation. Thus, it is interesting to ask whether a deviation from the BD states can leave any observational signatures in primordial non-Gaussianities to current/forthcoming observations. Such considerations have attracted a great deal of recent attentions, 
in which the non-Gaussianity generated from excited states during inflation is analyzed by several authors (for instance, see \cite{chen_observational_2007, holman_enhanced_2008, agullo_non-gaussianities_2011, agullo_stimulated_2011, ganc_calculating_2011, agullo_enhanced_2012, meerburg_signatures_2009, chen_folded_2010} and references therein). Depending on different types of inflationary models, the initial state effects may 
lead to enhancements in the associated non-Gaussianity in certain momentum configurations. These enhancements are sizable and could be well within detection in the forthcoming CMB experiments and large scale surveys. Even for the simplest single field slow-roll inflation, 
as shown in \cite{agullo_non-gaussianities_2011, ganc_calculating_2011}, such enhancements could occur in the squeezed 
configurations which involve very different scales.

On the other hand, recently released Planck data reported a hemispherical power asymmetry in the CMB fluctuations and provided an independent measurement on this anomaly \cite{ade_planck_2014, ade_planck_2016}, which was reported earlier in the Wilkinson microwave anisotropy probe data \cite{hansen_testing_2004, eriksen_n-point_2005, akrami_power_2014}. Such a power asymmetry can be modeled as a dipolar modulation of a statistically isotropic CMB sky in terms of temperature fluctuations in the direction ${\bf n}:$
\bqn
\frac{\delta T}{T}({\bf n}) = s({\bf n}) [1+ A \; {\bf n} \cdot \bf p],
\eqn
where $s({\bf n})$ is a statistically isotropic map, $A$ characterizes the amplitude of the dipolar asymmetry, and ${\bf p}$ is its direction. Then, 
the corresponding amplitude can be given by $A=0.066 \pm 0.021$ for the CMB power spectrum with $l \lesssim 64$ \cite{ade_planck_2014, ade_planck_2016}. This asymmetry, however, seems to vanish at larger scales, i.e. at scales $l \gtrsim 600,$ indicating a non-trivial scale-dependent dipole 
asymmetry \cite{hirata_constraints_2009, flender_small_2013}. This dipole asymmetry can be related to the primordial curvature perturbation 
power spectrum with a modulation as
\bqn
\mathcal{P}^{\rm mod}_{\mathcal{R}}(k, {\bf x}) = \mathcal{P}_{\mathcal{R}}(k) \left(1+ 2 A(k) \frac{{\bf p} \cdot {\bf x}}{x_{\rm cmb}}\right),
\eqn
where $x_{\rm cmb}$ is the comoving distance to the surface of the last scattering and $\mathcal{P}_{\mathcal{R}}(k)$ is the isotropic power spectrum. While such an asymmetry is very difficult to explain in a single-field slow-roll inflationary model, the challenge is therefore to find a mechanism being
able to generate it only at large scales. One of the interesting scenarios accounted for this power asymmetry is the 
Erickcek-Kamionkowski-Carroll (EKC) mechanism proposed in \cite{erickcek_superhorizon_2008, erickcek_hemispherical_2008}. This mechanism produces power asymmetry through a non-Gaussian coupling which requires having a growing amplitude in the squeezed limit between 
the observable scales and a large-amplitude perturbation mode at superhorizon scales. It is interesting to mention that both the superhorizon modes 
and enhanced non-Gaussianity in the squeezed limit could naturally arise from the effects of the non-BD excited states at the onset of the inflation \cite{schmidt_cosmic_2013} \footnote{It is worth noting that such superhorizon modes as well as the power spectrum modulation that account for this power asymmetry could arise from a lot of models. For examples see \cite{gao_can_2011, cai_cosmic_2014, liu_obtaining_2013} and references therein.}.

An example of generating a non-BD initial state arises naturally from the loop quantum cosmology, in which there is a bouncing phase prior to inflation. Remarkably, the dominating quantum geometry effects of LQC at the Planck scale provide a natural resolution of the standard big bang singularity (see \cite{ashtekar_loop_2011, ashtekar_loop_2015, barrau_some_2016, yang_alternative_2009} and references therein). In such a picture, the singularity is replaced by a finite non-zero universe, which eventually evolves  to the desired slow-roll inflation with very high probability \cite{singh_nonsingular_2006, mielczarek_inflation_2010, zhang_inflationary_2007, chen_loop_2015}.  The evolution of primordial perturbations during the pre-inflationary phase with different quantization approaches in LQC and their fingerprints on primordial power spectra have also been extensively studied both numerically and analytically \cite{agullo_quantum_2012, agullo_extension_2013, agullo_pre-inflationary_2013, bonga_inflation_2016, bonga_phenomenological_2016, bolliet_comparison_2015, schander_primordial_2016, bolliet_observational_2016, grain_perturbed_2016, mielczarek_gravitational_2008, linsefors_primordial_2013, mielczarek_possible_2010, zhu_universal_2016, zhu_pre-inflationary_2017, agullo_detailed_2015}.  The main characteristic of the associated effects is that the evolution of perturbations during the pre-inflationary phase produces  particles, and as a consequence the perturbations are no longer in the BD states at the onset of the slow-roll inflation, but instead in an excited state.

Therefore, it is important to find if the bouncing effects can provide any significant fingerprints on non-Gaussianities to produce observational effects. In fact, the non-Gaussianity and its effects on power asymmetry have been already studied in \cite{agullo_loop_2015} by using numerical calculations. The purpose of this paper is to provide a detailed and analytical study on non-Gaussianity in the primordial  curvature perturbations and their effects on the power asymmetry for a single field inflation in the framework of loop quantum cosmology, where there is a bouncing phase prior to inflation. Specifically, we focus on the primordial perturbations derived from the dressed metric approach in LQC, in which the bouncing effects are more important at large scales. We calculate in detail the analytical expression of the non-Gaussianity with quantum  effects. With these calculations, we show that the non-Gaussianity is negligible 
at observable scales, but enhanced when one of the coupled modes is at scales beyond our current Hubble horizon. We show explicitly, by using the EKC mechanism, that this large mode naturally leads to a significant modulation on observable scales which may provide an explanation of the
observed power asymmetry in the CMB.

We organize the rest of this paper as follows. In Sec. II, we provide a brief introduction to the evolution of both background and primordial 
perturbation modes during the bouncing phase and their effects on the primordial perturbation spectrum. Sec. III is devoted to analytical computations of the
non-Gaussianity of the primordial curvature perturbation. We show that the contributions due to quantum effects are of the same magnitude as the slow-roll parameter at observable scales, i.e., $f_{\rm NL} \sim \mathcal{O}(0.01)$. In Sec. IV, the non-Gausianity for a superhorizon mode coupled to observed scales is considered. We show that it can get enhanced and thus can lead to the observed power asymmetry in the CMB. 
Our main conclusions are presented in Sec. V.

\section{Non-BD state generated by pre-inflationary phase and its effects on primordial spectrum}
\renewcommand{\theequation}{2.\arabic{equation}}\setcounter{equation}{0}

A robust prediction of LQC is the occurrence of a non-singular bouncing phase, which removed the initial singularity in the early stage of the classical universe. In this section we present a brief introduction to both the evolution of the background with this pre-inflationary phase and its effects on primordial perturbations. 
For more details the reader is referred to \cite{zhu_pre-inflationary_2017, zhu_universal_2016}.

\subsection{Evolution of background during bouncing phase}

We first consider the evolution of the background for a flat Friedmann-Robertson-Walker (FRW) universe with a single scalar field $\phi$. In LQC, 
an effective semi-classical dynamics can be described by the modified Friedmann and Klein-Gordon equations,
\bqn
\lb{friedmann}
&&H^2=\frac{8\pi}{3m_{\text{Pl}}^2}\rho\left(1-\frac{\rho}{\rho_\text{c}}\right),\\
\lb{klein-gordon}
&&\ddot \phi +3 H \dot \phi +V_{,\phi}=0,
\eqn
where $H\equiv \dot a/a$ denotes the Hubble parameter and the dot represents the derivative with respect to the cosmic time $t$, $m_{\text{Pl}}=1/\sqrt{G}$ is the Planck mass, $\rho = \dot \phi^2/2+V(\phi)$ is the energy density of the universe with $V(\phi)$ being the potential of the scalar field, and $\rho_\text{c}$ is the critical energy density which represents the maximum value of the energy density in LQC and it is about $\rho_\text{c} \simeq 0.41 m_\text{Pl}^4$, as suggested in black hole entropy calculations.

Eq.~(\ref{friedmann}) indicates that the universe starts at $\rho=\rho_\text{c}$, where the energy density reaches its maximum value and the Hubble parameter becomes zero. The background evolution with a bouncing phase has been extensively studied, and one of the main results is that, immediately 
following the quantum bounce, a desired slow-roll inflation phase is almost inevitable \cite{ashtekar_loop_2011, singh_nonsingular_2006, mielczarek_inflation_2010, zhang_inflationary_2007, chen_loop_2015}.

To understand the background evolution, we first consider the initial conditions, which can be determined by specifying the values of the scalar field $\phi_{\rm B}$ and its velocity $\dot \phi_{\rm B}$. Among the whole $(\phi_{\rm B}, \dot \phi_{\rm B})$ space which satisfies $\dot \phi_{\rm B}^2/2 + V(\phi_{\rm B}) = \rho_{\rm c}$, we will focus on those parts in which the kinetic energy dominates at the beginning (or bounce) for two reasons. First, for kinetic energy dominated initial conditions, the background evolution during the bouncing phase is universal and can be solved analytically \cite{zhu_pre-inflationary_2017, zhu_universal_2016}. Second, a potential dominated bounce either is not able to produce the desired slow-roll inflation because of a lack of the initial kinetic energy or leads to a large e-folds of the slow-roll inflation. As pointed out in \cite{bonga_phenomenological_2016, zhu_pre-inflationary_2017, agullo_detailed_2015}, the latter possibility is out of our interest, because a huge amount of the slow-roll inflation washes out all the observational information of quantum effects and leads to the same perturbation spectrum as that in GR.

\begin{figure}
\includegraphics[width=8.1cm]{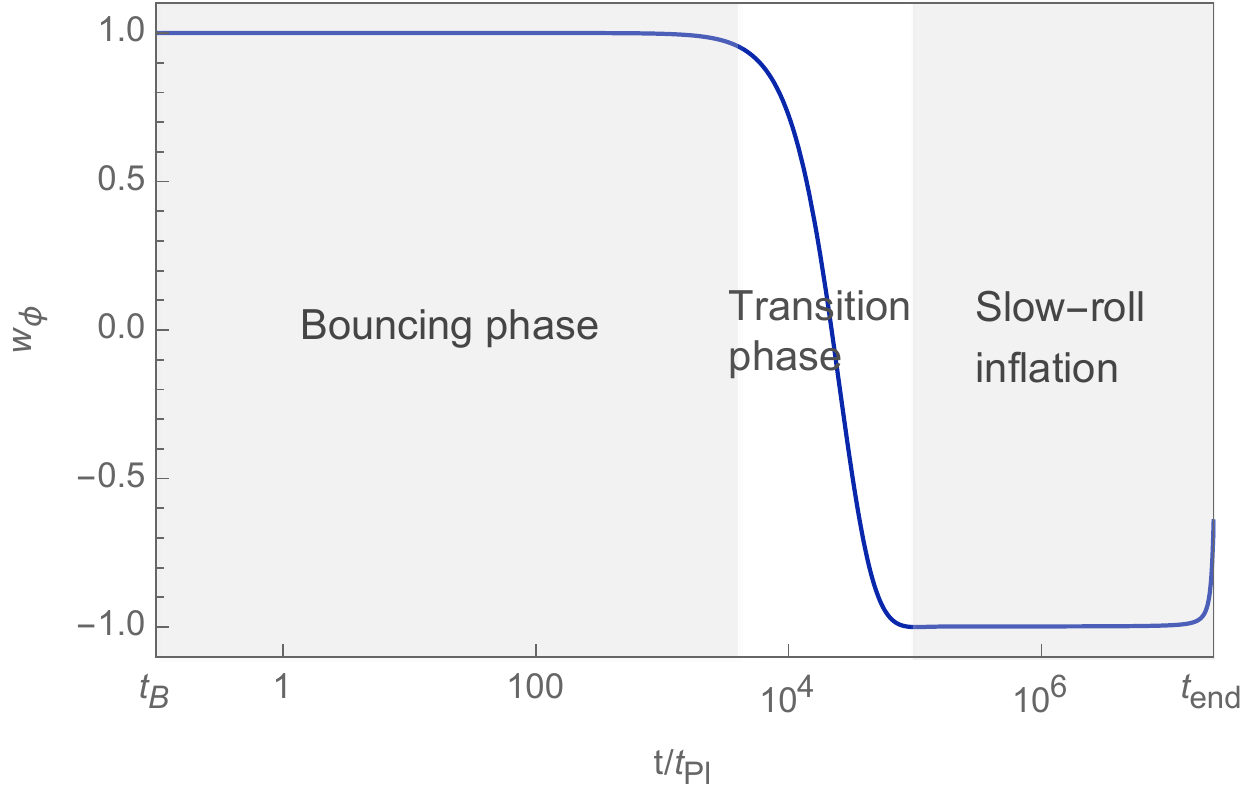}
\caption{Three different stages of the evolution of the universe from its initial state ($\rho = \rho_c$) until the end of the 
slow-roll inflation: {\em the bouncing,  transition, and  slow-roll inflation phases.} }
\label{wphi}
\end{figure}

When the kinetic energy of the inflation dominates in the beginning, it is found that the evolution of the universe can be 
divided universally into three stages prior to the reheating \cite{zhu_pre-inflationary_2017, zhu_universal_2016}:
{\em the bouncing, transition and slow-roll inflation}, as shown schematically in Fig.~\ref{wphi} for the evolution of the  equation of state $w(\phi) \equiv (\dot\phi^2 - 2V)/(\dot\phi^2 + 2V)$.  The universe is initially dominated by the kinetic energy ($w_\phi \simeq 1$), and then stays with $w_\phi \simeq 1$ during the whole bouncing phase. Then, the kinetic energy suddenly decreases and is soon dominated by the potential energy ($w_\phi\simeq -1$), whereby the slow-roll inflation takes over.
It is remarkable that the above division of the three different phases is universal, and does not depend on the inflationary models 
and initial conditions, as long as the universe was dominated by the kinetic energy of the inflation in the beginning.

For kinetic energy dominated initial states, the potential term in both Friedmann and Klein-Gordon equations can be ignored, and 
therefore we observe that
\bqn
\lb{fri}H^2 =\frac{1}{3M_{\text{Pl}}^2} \frac{1}{2}\dot \phi^2 \left(1-\frac{\dot \phi^2}{2\rho_{\text{c}}}\right),\\
\lb{kg}\ddot \phi+3 H \dot \phi =0.
\eqn
The above couple of equations can be solved analytically. From the Klein-Gordon equation (\ref{kg}) we obtain
\bqn\lb{phi_dot}
\dot \phi(t)= \pm \sqrt{2\rho_{\text{c}}} \left(\frac{a(t)}{a_\text{B}}\right)^{-3}.
\eqn
Here $\pm$ correspond to cases where $\dot \phi_\text{B}$ is positive or negative at the bounce, respectively. Substituting this into Eq. (\ref{fri}), we find
\bqn\lb{scalar_analytical}
a(t)=a_\text{B}\left(1+\gamma_\text{B} \frac{t^2}{t_{\text{Pl}}^2}\right)^{1/6},
\eqn
where $\gamma_\text{B}\equiv \frac{24\pi \rho_{\text{c}}}{m_{\text{Pl}}^4}$ is a dimensionless constant. With the analytical solution of $a(t)$, from Eq.~(\ref{phi_dot}) we acquire that
\bqn \lb{phi_sol}
\phi(t)=\phi_{\text{B}} \pm \frac{m_{\text{Pl}}}{2\sqrt{3\pi}}\text{arcsinh}{\left(\sqrt{\gamma_\text{B}}\frac{t}{t_{\text{Pl}}}\right)},
\eqn
and
\bqn \lb{phidot_sol}
\dot \phi(t)= \pm \frac{\sqrt{2\rho_\text{c}}}{(1+\gamma_\text{B} t^2/t_{\text{Pl}}^2)^{1/2}}\; .
\eqn

\subsection{Generations of the non-BD state during pre-inflationary phase}

Generally speaking, there are different ways to implement cosmological perturbations in the framework of LQC, 
namely the dressed metric approach \cite{agullo_quantum_2012, agullo_extension_2013, agullo_pre-inflationary_2013} 
and the deformed algebra approach \cite{grain_perturbed_2016}. In this paper, we tend to focus on the dressed metric approach introduced in \cite{agullo_quantum_2012, agullo_extension_2013, agullo_pre-inflationary_2013}. We only consider scalar perturbations
which can be described via the effective equation of motion,
\bqn\lb{eom}
\mu_k^{(s)}(\eta)'' + \left(k^2 - \frac{a''}{a}  + U^{(s)}(\eta) \right)\mu_k^{(s)}(\eta) =0,
\eqn
where $\mu_k^{(s)}(\eta) = z_s \mathcal{R}$ is related to the comoving curvature perturbations $\mathcal{R}$ with $z_s \equiv a\dot \phi/H$, and
\bqn
U^{(s)}(\eta) = a^2 (\mathfrak{f}^2 V(\phi) + 2 \mathscr{f} V_{, \phi}(\phi) + V_{, \phi\phi}(\phi) ),
\eqn
with $\mathfrak{f} \equiv \sqrt{24\pi G } \dot \phi /\sqrt{\rho}$. Above, a prime denotes the derivative with respect to the conformal time $\eta$, which  in terms of the cosmic time is given as
\bqn
\eta=\int_{t_\text{end}}^{t} \frac{dt'}{a(t')},
\eqn
where $t_{\rm end}$ is the time when the inflation ends. Thus correspondingly, we have
\bqn
\eta_{\rm end} =0, \;\;\; \eta_{\rm B} = \int_{t_{\rm end}}^{t_{\rm B}} \frac{dt}{a(t)},
\eqn
where $t_{\rm B}$ and $\eta_{\rm B}$ denote the cosmic and conformal times, respectively, at the bounce.

During the bouncing phase, as the kinetic energy dominates, $U^{(s)}(\eta)$ is negligible. Then for the purpose of studying the evolution of $\mu_k^{(s)}(\eta)$ during the bouncing phase, we consider the term
\bqn
\mathscr{V}(\eta) \equiv \frac{a''}{a} = a_{\rm B}^2 \frac{\gamma_{\rm B} m_{\rm Pl}^2 (1- \gamma_{\rm B} t^2 /t_{\rm Pl}^2)}{9(1+\gamma_{\rm B} t^2 /t_{\rm Pl}^2)^{5/3}}.
\eqn
Obviously, if Eq.~(\ref{eom}) can be considered as the Schr\"{o}dinger equation, then $\mathscr{V}(\eta)$ serves as an effective barrier during the bouncing phase. This potential can be approximated by a P\"{o}schl-Teller (PT) potential,
\bqn
\mathscr{V}^{\rm PT}(\eta) = \mathscr{V}_0 \cosh^{-1} \alpha (\eta -\eta_{\rm B}),
\eqn
 for which the analytical solution of the  corresponding  Schr\"{o}dinger equation is known, where $\mathscr{V}_0 =a_{\rm B}^2 \gamma_{\rm B} m_{\rm Pl}^2/3$ and $\alpha^2 = 2 a_{\rm B}^2 \gamma_{\rm B}m_{\rm Pl}^2$.

Introducing two new variables $x$ and $\mathcal{Y}(x)$ via
\bqn
x(\eta) &=& \frac{1}{1+e^{-2 \alpha (\eta - \eta_{\rm B})}}, \\
\mathcal{Y}(x) &=& \Big(x (1-x) \Big)^{ \frac{i k}{2 \alpha}} \mu_k^{(s)}(\eta),
\eqn
we find that Eq.~(\ref{eom}) reduces to
\bqn
x(1-x) \frac{d^2 \mathcal{Y}}{dx^2} + \left[a_3 - (a_1+a_2 +1 )x \right]\frac{d\mathcal{Y}}{dx} -a _1 a_2 \mathcal{Y}=0,\nb\\
\eqn
where
\bqn
a_1 &\equiv& \frac{1}{2} + \frac{\sqrt{\alpha^2-4 \mathscr{V}_0}}{2\alpha} - \frac{i k}{\alpha},\\
a_2 &\equiv& \frac{1}{2} -  \frac{\sqrt{\alpha^2-4 \mathscr{V}_0}}{2\alpha} - \frac{i k}{\alpha},\\
a_1 &\equiv&1- \frac{i k}{\alpha}.
\eqn
Note that in deriving the above equation, we have ignored the $U^{(s)}(\eta)$ term in Eq.~(\ref{eom}) as we are in the bouncing phase. The above equation is the standard hypergeometric equation, and its general solution is
\bqn
\mu_k^{(s)}(\eta) &=& a_k x^{ik/(2\alpha)} (1-x)^{- ik/(2\alpha)} \nb\\
&&~~~ \times \;_2 F_1(a_1-a_3+1, a_2 - a_3 +1, 2-a_3, x)\nb\\
&&~~~~ + b_k [x(1-x)]^{-ik/(2\alpha)} \; _2F_1(a_1,a_2,a_3,x).\nb\\
\eqn
Here $a_k$ and $b_k$ are two integration constants to be determined by the initial conditions. To impose them, let us first specify the initial time. A natural choice is right at the bounce, at which the initial state can be constructed as the fourth-order adiabatic vacuum \cite{agullo_pre-inflationary_2013}. While such constructions work well for large k,  ambiguity  however remains for modes with $k \lesssim k_{\rm B}$ \cite{agullo_pre-inflationary_2013}. Another choice that has been frequently used is the remote past \cite{bolliet_comparison_2015, schander_primordial_2016}. In this paper we adopt the latter and in \cite{zhu_pre-inflationary_2017}, we have proved that these two different choices actually lead to the same results    \cite{zhu_pre-inflationary_2017}. As pointed out in \cite{bolliet_comparison_2015, schander_primordial_2016, zhu_pre-inflationary_2017}, during the contracting phase right before the bounce, all the relevant perturbation modes are well inside the horizon, so that we can naturally choose the BD vacuum state as the initial conditions. Then the coefficients $a_k$ and $b_k$ can be determined as
\bqn
a_k =0 ,\;\;\; b_k = \frac{e^{i k \eta_{\rm B}}}{\sqrt{2k}}.
\eqn

After the quantum bounce, the universe will soon turn into the slow-roll inflation, and the mode function is given by
\bqn\lb{mode}
\mu_k^{(s)}(\eta) = \alpha_k \frac{e^{- i k \eta}}{\sqrt{2k}} \left(1- \frac{i}{k\eta}\right) + \beta_k  \frac{e^{ i k \eta}}{\sqrt{2k}} \left(1+ \frac{i}{k\eta}\right).\nb\\
\eqn
Matching this solution with that during the bouncing phase, it is easy to show that at the onset of the slow-roll inflation, the perturbation modes take the form
\bqn
\mu_k^{(s)}(\eta) =  \alpha_k \frac{e^{- i k \eta}}{\sqrt{2k}}  + \beta_k  \frac{e^{ i k \eta}}{\sqrt{2k}},
\eqn
with
\bqn
\alpha_k &=& \frac{\Gamma(a_3) \Gamma(a_3-a_1 -a_3)}{\Gamma(a_3-a_1) \Gamma(a_3-a_2)} e^{2 i k \eta_{\rm B}},\\
\beta_k &=& \frac{\Gamma(a_3) \Gamma(a_1+a_2-a_3)}{\Gamma(a_1) \Gamma(a_2)}.
\eqn
In GR, one in general imposes the BD vacuum state when the modes are inside the Hubble horizon, i.e., $\alpha_k=1,\; \beta_k=0$. However, we show clearly that if there is a bouncing phase prior to the slow-roll inflation, $\beta_k$ now does not vanish generically. This leads to modifications at the onset of the slow-roll inflation on the standard nearly scale invariant power spectrum,
\bqn\lb{pw}
\mathcal{P}_{\mathcal{R}}(k) = |\alpha_k+\beta_k|^2 \frac{H^2}{8 \pi^2 m_{\rm Pl}^2 \epsilon_1^2},
\eqn
where
\bqn
&&|\alpha_k+\beta_k|^2 \nb\\
&& ~~ =1+ \left[1+\cos\left(\frac{\pi}{\sqrt{3}}\right)\right]{\rm csch}^2\left(\frac{\pi k}{\sqrt{6} k_{\rm B}}\right)\nb\\
&& ~~~~ + \sqrt{2} \cos\left(\frac{\pi}{2\sqrt{3}}\right)\sqrt{\cos\left(\frac{\pi}{\sqrt{3}} \right)+ \cosh\left(\frac{2 \pi k}{\sqrt{6} k_{\rm B}}\right)}\nb\\
&&~~~~~  \times {\rm csch}^2\left(\frac{\pi k}{\sqrt{6} k_{\rm B}}\right) \cos(2 k_{\rm B} \eta_{\rm B}+\varphi_k),
\eqn
with $\epsilon_1 \equiv -  \dot H/H^2$ and
\bqn
\varphi_k \equiv \arctan \Bigg\{\frac{\Im [\Gamma(a_1) \Gamma(a_2) \Gamma^2(a_3-a_1-a_2)]}{\Re [\Gamma(a_1) \Gamma(a_2) \Gamma^2(a_3-a_1-a_2)]}\Bigg\}.\nb
\eqn

\subsection{Observational range of pre-inflationary effects}

A bouncing phase prior to inflation leads to an excited state at the onset of the slow-roll inflation. This excited state in turn produces scale 
dependent features in the primordial perturbation spectrum, as described in (\ref{pw}). These effects are encoded in the Bogoliubov 
coefficients $\alpha_k$ and $\beta_k$, and essentially depend on the parameter $k_{\rm B}$. Thus they
represent a characteristic feature of the LQC.

The quantum effects on the primordial perturbation spectrum can be constrained by recent observational data. Recently, by 
utilizing the Planck 2015 data, we found  that the parameter 
$k_{\rm B}/a_0$ is constrained by Planck TT + lowP (Planck TT, TE, EE + lowP) to \cite{zhu_universal_2016, zhu_pre-inflationary_2017}, 
\bqn
\frac{k_{\rm B}}{a_0} < 3.12 \times 10^{-4} {\rm Mpc}^{-1} (3.05 \times 10^{-4}),
\eqn
at 95\% C.L. and constrained by Planck TT + lowP + tensor (Planck TT, TE, EE + lowP + tensor) to
\bqn
\frac{k_{\rm B}}{a_0} < 3.14 \times 10^{-4} {\rm Mpc}^{-1} (3.14 \times 10^{-4}),
\eqn
at 95\% C.L.. Considering the fact that the observable modes lie in the range of $(k_{\rm min}/a_0, \; k_{\rm max}/a_0)$ with $k_{\rm min}/a_0 = H_0 \simeq 3.3 \times 10^{-4} {\rm Mpc}^{-1}$ and $k_{\rm max}/a_0 \simeq 1 {\rm Mpc}^{-1}$, it is easy to infer from the above constraints that
 \bqn\lb{kbkmin}
 \frac{k_{\rm B}}{a_0} \lesssim \frac{k_{\rm min}}{a_0}.
 \eqn
 This indicates that the modes with $k \lesssim k_{\rm B}$ are outside of the current Hubble horizon and only modes with $k \gtrsim k_{\rm B}$ could lie in the observational range.

Based on the above constraints, the bouncing effects in the Bogoliubov coefficients $\alpha_k$, $\beta_k$ and the corresponding power spectrum can be divided into three different regions,  the enhanced region, observable region, and suppressed region.

The enhanced region corresponds to the modes with $k \ll k_{\rm B}$. For these modes, it is convenient to expand the Bogoliubov coefficients $\alpha_k$ and $\beta_k$ about $k/k_{\rm B}$ as
\bqn
|\alpha_k| &\simeq& \frac{2\sqrt{3}}{\pi} \cos\left(\frac{\pi}{2\sqrt{3}}\right) \frac{k_{\rm B}}{k} \gg 1,\\
|\beta_k| &\simeq &\frac{\sqrt{6}}{\pi} \cos\left(\frac{\pi}{2\sqrt{3}}\right) \frac{k_{\rm B}}{k} \gg 1.
\eqn
This implies that the state of perturbation modes dramatically deviate from the BD state and the corresponding effects dominate on these scales. As a result, the primordial perturbation spectrum is significantly enhanced due to the existence of these excited states.

The suppressed region corresponds to another limit with $k \gg k_{\rm B}$. For these modes, the two Bogoliubov coefficients can be approximately expressed as
\bqn
|\alpha_k | &\simeq& 1 + \left[1 + \cos\left(\frac{\pi}{\sqrt{3}}\right)\right] e^{- \frac{2 \pi}{ \sqrt{6}} \frac{k}{k_{\rm B}}} ,\\
|\beta_k| & \simeq & 2 \cos\left(\frac{\pi}{2\sqrt{3}}\right) e^{- \frac{\pi}{\sqrt{6}} \frac{k}{k_{\rm B}}} .
\eqn
In this limit, the relevant modes reduce to the standard BD state, ie., $\alpha_k \to 1$ and $\beta_k \to 0$. Thus the bouncing effects are suppressed by the factor $e^{- \frac{\pi}{\sqrt{6}} \frac{k}{k_{\rm B}}}$ and the resulting power spectrum then reduces to the standard one.

The most interesting region, which relates to the observable region with significant bouncing effects, corresponds to the modes with $k \gtrsim k_{\rm B}$. To estimate the bouncing effects, it is convenient to calculate the Bogoliubov coefficients $\alpha_k$ and $\beta_k$ at $k=k_{\rm B}$ as
\bqn
|\alpha_k| &=& \text{csch}\left(\frac{\pi }{\sqrt{6}}\right) \sqrt{\frac{1}{2} \cos \left(\frac{\pi }{\sqrt{3}}\right)+ \frac{1}{2}\cosh \left(\sqrt{\frac{2}{3}} \pi \right)}, \nb\\
|\beta_k| &=& \cos \left(\frac{\pi }{2 \sqrt{3}}\right) \text{csch}\left(\frac{\pi }{\sqrt{6}}\right).
\eqn
Approximately we find $|\alpha_k| \simeq 1.06$ and $|\beta_k| \simeq 0.37$ when $k=k_{\rm B}$. These results shall play an essential role in the estimation of the non-Gaussianity in the observational range in the next section.

\section{Non-Gaussianity in Bispectrum from quantum bounce effects}
\renewcommand{\theequation}{3.\arabic{equation}}\setcounter{equation}{0}

In this section, we begin studying the non-Gaussianity of the comoving curvature perturbation $\mathcal{R}_k.$ 
It is important to note that the non-Gaussianity with quantum gravitational effects has been calculated numerically 
in \cite{agullo_loop_2015}. In this section, we shall follow the procedure in \cite{agullo_loop_2015} while employing 
our analytical solution obtained in the previous section.

In general, the non-Gaussianity of the comoving curvature perturbations $\mathcal{R}_k$ is characterized by 
the bispectrum $B_{\mathcal{R}}(k_1, k_2, k_3)$, which is defined in terms of a three-point correlation function as
\bqn
\langle \mathcal{R}_{\bf k_1} \mathcal{R}_{\bf k_2} \mathcal{R}_{\bf k_3} \rangle = (2\pi)^3 \delta^3 ({ \bf k_1}+ {\bf k_2} +{\bf k_3}) B_{\mathcal{R}}(k_1, k_2, k_3),\nb\\
\eqn
where the three point function for the comoving curvature perturbation can be calculated via
\bqn\lb{nG_integral}
\langle \mathcal{R}_{\bf k_1}  \mathcal{R}_{\bf k_2} \mathcal{R}_{\bf k_3} \rangle &\simeq& - i \langle \left[ \mathcal{R}_{\bf k_1}  \mathcal{R}_{\bf k_2}  \mathcal{R}_{\bf k_3}, \int_{t_0}^t dt' H_3(t') \right] \rangle, \nb\\
\eqn
where $H_3(t)$ denotes the cubic Hamiltonian of the comoving curvature perturbation $\mathcal{R}({\bf x},t)$ in the metric and it is expressed as \cite{wang_inflation_2014}
\bqn
H_3(t) &=& - \int d^3 x \Bigg\{a^3 \epsilon_1^2 \mathcal{R} \mathcal{\dot R}^2 + a \epsilon_1^2 \mathcal{R} (\partial \mathcal{R})^2  \nb\\
&&~~~~~~~~~~~~ - 2 a^3 \epsilon_1^2 \mathcal{\dot R} \partial \mathcal{R} \partial (\partial^{-2} \mathcal{\dot R}) \nb\\
&&~~~~~~~~~~~~ +\partial_t \left(- \frac{\epsilon_1 \epsilon_2}{2} a^3 \mathcal{R}^2 \mathcal{\dot R}\right)\Bigg\}.
\eqn
Performing the integral in Eq.~(\ref{nG_integral}) for the mode function in Eq.~(\ref{mode}) with a general $\alpha_k$ and $\beta_k $,  one obtains \footnote{The non-Gaussianity with non-BD states has been also calculated in \cite{holman_enhanced_2008, agullo_non-gaussianities_2011, agullo_stimulated_2011, ganc_calculating_2011, agullo_enhanced_2012, meerburg_signatures_2009, chen_folded_2010}}
\begin{widetext}
\bqn
&&B_\mathcal{R}(k_1,k_2,k_3) \nb\\
&&~~ = (2\pi)^4  \left(\frac{H^2}{8\pi^2 M_{\rm Pl}^2 \epsilon_1} \right)^2 \frac{ (\alpha_{k_1}+\beta_{k_1})(\alpha_{k_2}+\beta_{k_2})(\alpha_{k_3}+\beta_{k_3})}{8 k_1^2 k_2^2 k_3^2}\nb\\
&&~~ ~~~~ \times \Bigg\{  (\epsilon_1+\epsilon_2)\frac{k_1^2+k_2^2+k_3^2}{2 k_1 k_2 k_3} \Bigg[ K ( \alpha_{k_1}^*  \alpha_{k_2}^*  \alpha_{k_3}^*- \beta_{k_1}^*  \beta_{k_2}^*  \beta_{k_3}^* )  + K_3(\alpha_{k_1}^*  \alpha_{k_2}^*  \beta_{k_3}^* - \beta_{k_1}^*  \beta_{k_2}^*  \alpha_{k_3}^*)  \nb\\
&&~~ ~~~~ ~~~~~~~~~~~~~~~~~~~~~~~~~~~~~~~~~~~ +K_2( \alpha_{k_1}^*  \beta_{k_2}^*  \alpha_{k_3}^* - \beta_{k_1}^*  \alpha_{k_2}^*  \beta_{k_3}^* ) - K_1(\alpha_{k_1}^*  \beta_{k_2}^*  \beta_{k_3}^* - \beta_{k_1}^*  \alpha_{k_2}^*  \alpha_{k_3}^*)
 \Bigg]\nb\\
&&~~ ~~~~ ~~~~~ + \Bigg[ \left(\frac{4 \left(k_1^2 k_2^2+k_3^2 k_2^2+k_1^2 k_3^2\right)-\left(k_1 k_2+k_3 k_2+k_1 k_3\right) \left(k_1^2+k_2^2+k_3^2\right)-\left(k_1^4+k_2^4+k_3^4\right)}{K k_1 k_2 k_3}+1\right)\epsilon _1 \nb\\
&&~~~~ ~~ ~~~~~~~~ ~~ - \frac{k_2 k_1^2+k_3 k_1^2+k_2^2 k_1+k_3^2 k_1+k_2 k_3^2+k_2^2 k_3}{2 k_1 k_2 k_3} \epsilon_2 \Bigg] \Big(\alpha_{k_1}^*  \alpha_{k_2}^*  \alpha_{k_3}^* (1-e^{i K\eta_0})- \beta_{k_1}^*  \beta_{k_2}^*  \beta_{k_3}^* (1-e^{- i K\eta_0})\Big)\nb\\
&&~~ ~~~~ ~~~~~ + \Bigg[ \left(\frac{4 \left(k_1^2 k_2^2+k_3^2 k_2^2+k_1^2 k_3^2\right)-\left(k_1 k_2-k_3 k_2-k_1 k_3\right) \left(k_1^2+k_2^2+k_3^2\right)-\left(k_1^4+k_2^4+k_3^4\right)}{K_3 k_1 k_2 k_3}-1\right)\epsilon _1 \nb\\
&&~~~~ ~~ ~~~~~~~~ ~~ - \frac{k_2 k_1^2-k_3 k_1^2+k_2^2 k_1+k_3^2 k_1+k_2 k_3^2-k_2^2 k_3}{2 k_1 k_2 k_3} \epsilon_2 \Bigg] \Big(\alpha_{k_1}^*  \alpha_{k_2}^*  \beta_{k_3}^* (1-e^{i K_3\eta_0})- \beta_{k_1}^*  \beta_{k_2}^*  \alpha_{k_3}^* (1-e^{- i K_3\eta_0})\Big)\nb\\
&&~~ ~~~~ ~~~~~ + \Bigg[ \left(\frac{4 \left(k_1^2 k_2^2+k_3^2 k_2^2+k_1^2 k_3^2\right)-\left(k_1 k_3-k_1 k_2-k_3 k_2\right) \left(k_1^2+k_2^2+k_3^2\right)-\left(k_1^4+k_2^4+k_3^4\right)}{K_2 k_1 k_2 k_3}-1\right)\epsilon _1 \nb\\
&&~~~~ ~~ ~~~~~~~~ ~~ - \frac{- k_2 k_1^2+k_3 k_1^2+k_2^2 k_1+k_3^2 k_1-k_2 k_3^2+k_2^2 k_3}{2 k_1 k_2 k_3} \epsilon_2 \Bigg] \Big(\alpha_{k_1}^*  \beta_{k_2}^*  \alpha_{k_3}^* (1-e^{i K_2\eta_0})- \beta_{k_1}^*  \alpha_{k_2}^*  \beta_{k_3}^* (1-e^{- i K_2\eta_0})\Big)\nb\\
&&~~ ~~~~ ~~~~~ - \Bigg[ \left(\frac{4 \left(k_1^2 k_2^2+k_3^2 k_2^2+k_1^2 k_3^2\right)-\left(k_3 k_2-k_1 k_3-k_1 k_2\right) \left(k_1^2+k_2^2+k_3^2\right)-\left(k_1^4+k_2^4+k_3^4\right)}{K_1 k_1 k_2 k_3}-1\right)\epsilon _1 \nb\\
&&~~~~ ~~ ~~~~~~~~ ~~ + \frac{ k_2 k_1^2+k_3 k_1^2-k_2^2 k_1-k_3^2 k_1+k_2 k_3^2+k_2^2 k_3}{2 k_1 k_2 k_3} \epsilon_2 \Bigg] \Big(\alpha_{k_1}^*  \beta_{k_2}^*  \beta_{k_3}^* (1-e^{-i K_1\eta_0})- \beta_{k_1}^*  \alpha_{k_2}^*  \alpha_{k_3}^* (1-e^{ i K_1\eta_0})\Big)\Bigg\}\nb\\
&&~~~ +  {\rm c.c.},
\eqn
\end{widetext}
where $K \equiv k_1 +k_2 +k_3$ and $K_i \equiv K-2 k_i$ with $i =1, 2, 3$. Then to describe the amplitude of the bispectrum, it is also convenient to define the amplitude $f_{\rm NL}$ as
\bqn
f_{\rm NL} \equiv  \frac{\frac{5}{6} B_{\mathcal{R}} (k_1, k_2, k_3)}{P_{\mathcal{R}}(k_1) P_{\mathcal{R}}(k_2) + P_{\mathcal{R}}(k_1) P_{\mathcal{R}}(k_3) + P_{\mathcal{R}}(k_2) P_{\mathcal{R}}(k_3) },\nb\\
\eqn
where
\bqn
P_{\mathcal{R}}(k) \equiv |\alpha_k+\beta_k|^2 \frac{H^2}{8 \pi^2  M_{\rm Pl}^2 \epsilon_1} \frac{2\pi^2}{k^3}.
\eqn

In general, the amplitude $f_{\rm NL} \sim \mathcal{O}(\epsilon_1, \epsilon_2) $, i.e., it is suppressed by the 
slow-roll parameters. As $\alpha_k=1$ and $\beta_k=0,$ it is readily  observed that $f_{\rm NL} \sim 0.01$. Thus it is too small 
to be detectable. When $\beta_k \neq 0,$ there are two specific momentum configurations, the folded and squeezed configurations,  
that are expected to produce large non-Gaussianities.

\subsection{Folded configuration}

For the folded limit ($k_1= k_2=\frac{k_3}{2}$), we have $K_3 \to 0$. Thus one expects that the bispectrum is dominated by 
the term $\frac{1- e^{\pm i K_3 \eta_0}}{K_3}.$ In this case we adopt the limitation
\bqn
\lim_{K_3 \to 0} \frac{1-e^{\pm i K_3 \eta_0}}{K_3} = \mp i \eta_0.
\eqn
It follows immediately that
\bqn
f_{\rm NL}^{\rm folded} &\simeq & \epsilon_1  \times \mathcal{O}(1) \times  k_3 \eta_0.
\eqn
Considering the fact that the mode $k_3$ should be well within the horizon at the onset of the slow-roll inflation (or right after the bounce), we have
\bqn
- k_3 \eta_0  \gg 1.
\eqn
Thus, we may anticipate that $k_3 \eta_0$ should make significant contributions to the non-Gaussianity. However, 
as pointed out in \cite{holman_enhanced_2008}, when we consider the two-dimensional projections of the three point function into the CMB, 
this factor $k_3 \eta_0$ will essentially disappear. So this result is not enhanced but will be of the same order as that from the local 
redefinitions and the two-point function.

\subsection{Squeezed configuration}

For the squeezed configuration ($ k_2 \simeq k_3 \gg k_1$), we have
\bqn
\frac{1}{K_2} \to \frac{1}{k_1}, \;\; \frac{1}{K_3} \to \frac{1}{k_1}.
\eqn
Considering that $k_1$ is small enough compared to the other two modes in this case, we find
\bqn\lb{fnl}
f_{\rm NL}^{\rm squeezed} &\simeq& \frac{10}{3} \epsilon_1 \left[\frac{k_3}{k_1} + \frac{9}{4} \frac{k_1}{k_3} +\mathcal{O}(k_1^2/k_3^2)\right] \nb\\
&& \times {\rm Re} \Big[ \frac{(\alpha_{k_1} + \beta_{k_1})(\alpha_{k_3} + \beta_{k_3})^2 }{|\alpha_{k_1} +\beta_{k_1}|^2 |\alpha_{k_3}+\beta_{k_3}|^2}\nb\\
&&~~~~~~ \times  \Big(\alpha_{k_1}^*  \alpha_{k_3}^*  \beta_{k_3}^* - \beta_{k_1}^*  \beta_{k_3}^*  \alpha_{k_3}^*   \Big) \Big].\nb\\
\eqn
In the observational range,  $k_3/k_1$ could be as large as $k_{\rm max}/k_{\rm min} \sim 10^{4}$. Thus, 
 in general one expects in the squeezed limit the non-Gaussianity gets enhanced due to the excited state, i.e., $\beta_k \neq 0$.  However, whether the non-Gaussianity gets enhanced or not also depends on the magnitude of the term in the second line of Eq.~(\ref{fnl}). In order to see the quantum effects in this limit, let us study this term in detail.

We take the large scale $k_1$ as the largest scale we could observe today, thus we can identify it as the typical scale of quantum effects $k_{\rm B}$. With this consideration we find
\bqn
&& \frac{(\alpha_{k_{\rm B}} + \beta_{k_{\rm B}})(\alpha_{k_3} + \beta_{k_3})^2}{|\alpha_{k_{\rm B}} + \beta_{k_{\rm B}}|^2 |\alpha_{k_3}+\beta_{k_3}|^2}  \Big(\alpha_{k_{\rm B}}^*  \alpha_{k_3}^*  \beta_{k_3}^* - \beta_{k_{\rm B}}^*  \beta_{k_3}^*  \alpha_{k_3}^* \Big) \nb\\
&&~~~~~~~~ \simeq \frac{\alpha_{k_{\rm B}}^*-\beta_{k_{\rm B}}^*}{\alpha_{k_{\rm B}}^*+\beta_{k_{\rm B}}^*} \alpha_{k_3}^* \beta_{k_3}^* + \mathcal{O}(\beta_{k_3}^2),
\eqn
where we assumed that $k_{3} \gg k_{\rm B}$ thus $|\beta_{\rm k_3}| \ll 1$. Using the approximate form
\bqn
\beta_{k_3}^* \simeq i 2 \cos{\left( \frac{\pi}{2\sqrt{3}}\right)} e^{- \frac{\pi}{\sqrt{6}} \frac{k_3}{k_{\rm B}}} + \mathcal{O}(e^{- \frac{2\pi}{\sqrt{6}} \frac{k_3}{k_{\rm B}}}),
\eqn
we find
\bqn
f_{\rm NL}^{\rm squeezed} &\lesssim& \frac{20}{3} \cos{\left( \frac{\pi}{2\sqrt{3}}\right)} \epsilon_1\times \frac{k_3}{k_{\rm B}} e^{-\frac{\pi}{\sqrt{6}} \frac{k_3}{k_{\rm B}}}\nb\\
&& \times {\rm Im} \left(\frac{\alpha_{k_{\rm B}}^*-\beta_{k_{\rm B}}^*}{\alpha_{k_{\rm B}}^*+\beta_{k_{\rm B}}^*} \alpha^*_{k_3} \right),\nb\\
\eqn
where the second line of the above equation is of the order of $\mathcal{O}(1)$ and
\bqn
\frac{k_3}{k_{\rm B}} e^{- \frac{\pi}{\sqrt{6}} \frac{k_3}{k_{\rm B}}} \ll e^{- \frac{\pi}{\sqrt{6}}} \simeq 0.28.
\eqn
Therefore, the above expression shows that the quantum effects in the squeezed limit are strongly suppressed by the factor $e^{-\frac{\pi}{\sqrt{6}} \frac{k_3}{k_{\rm B}}}$,  even though $k_3/k_{\rm B} \gg 1$. This leads to one of the main conclusion of this paper:  the quantum gravitational effects on the non-Gaussianity only contribute to the same order as that in a single field slow-roll inflation in the observable range. It is far beyond our current or forthcoming detections. It is worth to mention that our result presented is in a good agreement with the discussion in \cite{agullo_loop_2015}.

\section{Enhanced Quantum effects at superhorizon scales and Power Asymmetry}
\renewcommand{\theequation}{4.\arabic{equation}}\setcounter{equation}{0}

We now turn to the non-Gaussianity at scales beyond our current Hubble horizon and its possible effects on the observational power spectrum. According to the mechanism proposed in \cite{erickcek_superhorizon_2008, erickcek_hemispherical_2008}, the enhanced non-Gaussianity in the squeezed limit for coupling between superhorizon mode and observable scales could lead to a large modulation of the primordial spectrum.
Thus providing an origin for the power asymmetry. In this section, we follow this idea to consider enhanced quantum effects at superhorizon scales on the non-Gaussianity and their impacts on the power asymmetry of a CMB spectrum.

To illustrate the scenario with quantum effects, let us first consider the non-Gaussianity in the squeezed limit for a coupling between a superhorizon mode with $k_1=k_{\rm sh} \ll k_{\rm B}$ and two large scale modes with $k_2=k_3=k \gtrsim k_{\rm B}$ in the observational range. From Eq.~(\ref{fnl}), we have
\bqn
f_{\rm NL}^{\rm squeezed} &\simeq&  \frac{10}{3} \epsilon_1 \frac{k}{k_{\rm sh}} {\rm Re} \Big[ \frac{(\alpha_{k_{\rm sh}} + \beta_{k_{\rm sh}})(\alpha_{k} + \beta_{k})^2}{|\alpha_{k} +\beta_k|^2|\alpha_{k_{\rm sh}} + \beta_{k_{\rm sh}}|^2} \nb\\
&&~~~~~~ \times  \Big(\alpha_{k_{\rm sh}}^*  \alpha_{k}^*  \beta_{k}^* - \beta_{k_{\rm sh}}^*  \beta_{k}^*  \alpha_{k}^*   \Big) \Big].
\eqn
More explicitly we need to calculate effects due to the Bogolibov coefficients $\alpha_{k_{\rm sh}}$ and $\beta_{k_{\rm sh}}$, since they are expected to be enhanced as we show in Sec. II. C. Performing the Taylor expansions about $k_{\rm sh}/k_{\rm B}$, we acquire that
\bqn \lb{fnl_squeezed}
f_{\rm NL}^{\rm squeezed}(k) &\simeq& \epsilon_1  \frac{20\sqrt{6}}{3}  \frac{\cos \left(\frac{\pi }{2 \sqrt{3}}\right) \text{csch}\left(\frac{\pi  k}{\sqrt{6} k_{\rm B}}\right)}{H_{- \frac{1}{2}-\frac{\sqrt{3}}{6} }+H_{-\frac{1}{2} + \frac{\sqrt{3} }{6} }}  \frac{k k_{\rm B}}{k^2_{\rm sh}}  \nb\\
&&\times {\rm Im}\left( \frac{\alpha_{k} + \beta_{k}}{\alpha_{k}^* +\beta_k^*} \alpha_{k}^*  \right),
\eqn
where $H_{\nu}$ represents the harmonic number. Considering
\bqn
{\rm Im}\left( \frac{\alpha_{k} + \beta_{k}}{\alpha_{k}^* +\beta_k^*} \alpha_{k}^*  \right) \sim \mathcal{O}(1),
\eqn
and $k_{\rm sh} \ll k_{\rm B} \lesssim k$, it is obvious that $f_{\rm NL}^{\rm squeezed}$ is scale dependent (depends on $k$). It is enhanced dramatically by the factor $k k_{\rm B}/k_{\rm sh}^2$ at large scales,  but suppressed by ${\rm csch}{(\pi k /(\sqrt{6} k_{\rm B}))}$ at small scales. The presence of the superhorizon mode $k_{\rm sh}$ and its coupling to observational scales are the key ingredients to make $f_{\rm NL}$ enhanced.

The non-Gaussianity in the squeezed limit given in Eq.~(\ref{fnl_squeezed}) can produce modulation on the primordial curvature power spectrum. In particular, due to the EKC mechanism \cite{erickcek_superhorizon_2008, erickcek_hemispherical_2008}, the superhorizon mode could bring modifications at observational scales, which is expected as an approximately linear function of positions. This, naturally, could provide an explanation for the observed power asymmetry in the CMB spectrum. According to the analysis given in \cite{lyth_cmb_2013, namjoo_hemispherical_2013}, the relation between the power asymmetry and non-Gaussianity is given by
\bqn\lb{power_asymmetry}
A(k) = \frac{6}{5}|f_{\rm NL}^{\rm squeezed}(k) | k_{\rm sh}   x_{\rm cmb} \mathcal{P}^{1/2}_{\mathcal{R}}(k_{\rm sh}).
\eqn
This equation is also known as the consistency condition, relating the amplitude of power asymmetry to the amplitude of the non-Gaussianity in the squeezed limit. The remarkable feature of the non-Gaussianity produced by the quantum effects in the squeezed limit here is that it is sensitive to different scales. The amplitude of the non-Gaussianity in the squeezed limit, which is enhanced at large scales but suppressed at small scales (as we showed above),  could provide a physical explanation for the observational fact that the power asymmetry is only significant on the cosmological scale but becomes small at the ${\rm Mpc}$ scale.

Specifically, let us consider the primordial curvature perturbation spectrum at superhorizon scales. According to Eq.~(\ref{pw}), it is approximately given by
\bqn
\mathcal{P}_{\mathcal{R}}(k_{\rm sh}) \simeq \frac{6(1+\cos{(\pi/\sqrt{3})})}{\pi^2} \frac{k_{\rm B}^2}{k_{\rm sh}^2} \frac{H^2}{8 \pi^2 m_{\rm Pl}^2 \epsilon_1}.
\eqn
Then substituting this equation into Eq.~(\ref{power_asymmetry}) and using Eq.~(\ref{fnl_squeezed}) we obtain
\bqn
A(k) &\simeq&- \epsilon_1  \frac{48 \sqrt{1+\cos{(\pi/\sqrt{3})}} \cos \left(\frac{\pi }{2 \sqrt{3}}\right)}{ \pi \left(H_{- \frac{1}{2}-\frac{\sqrt{3}}{6} }+H_{-\frac{1}{2} + \frac{\sqrt{3} }{6} }\right)} \nb\\
&& \times \text{csch}\left(\frac{\pi  k}{\sqrt{6} k_{\rm B}}\right) \frac{k_{\rm B}^2}{k^2_{\rm sh}} \frac{k x_{\rm cmb}}{a_0} \sqrt{\frac{H^2}{8 \pi^2 m_{\rm Pl}^2 \epsilon_1}}.
\eqn
The Planck 2015 data showed  that $\epsilon_1 < 0.0068$ at $95\%$ C.L. and the best-fit value of $H^2/(8 \pi^2 m_{\rm Pl}^2 \epsilon)$ reads $ 2.2 \times 10^{-9}$. Substituting these values into $A(k)$ yields an upper bound,
\bqn
A(k) \lesssim 5.4\times 10^{-7} \text{csch}\left(\frac{\pi  k}{\sqrt{6} k_{\rm min}}\right) \frac{k_{\rm min}^2}{k^2_{\rm sh}} \frac{k x_{\rm cmb}}{a_0},
\eqn
where the constraint $k_{\rm B}< k_{\rm min}$ from Eq.~(\ref{kbkmin}) has been also applied. Typically, the magnitude of $A(k)$ at different scales is a function of $k_{\rm sh}$. Namely, as $A(k) \sim 0.066$ at large scales $(k/a_0)^{-1} \gtrsim (a_0/k_{\rm min} )^{-1} \sim 3 \; {\rm Gpc}$, we expect
\bqn
\frac{k_{\rm sh}}{a_0} \lesssim 1.2 \times 10^{-6} {\rm Mpc}^{-1},
\eqn
where we have used $x_{\rm cmb}=14 \; {\rm Gpc}$. At small scales, as the Bogoliubov coefficient $\beta_k$ is highly suppressed, the non-Gaussianity amplitude $f_{\rm NL}$ reduces to the usual magnitude with $\beta_k =0$, i.e., $f_{\rm NL} \sim n_s-1$. Thus, at small scales, the power asymmetry is small, which is consistent with the constraint from quasars \cite{hirata_constraints_2009}.

In conclusion, we have shown analytically that the quantum gravitational effects can produce a large-amplitude superhorizon mode which could account for the observed power asymmetry in the CMB. This is in agreement with the results obtained from numerical calculations in \cite{agullo_loop_2015}, in which the calculations are limited to specific initial conditions and the potential of the scalar field. In addition, we also would like to emphasize that, although the quantum effects produce power asymmetry which is consistent with observation, their effects in both the primordial perturbation spectrum and non-Gaussianity are well within current observational constraints.

\section{Conclusions}

In this paper,  we have provided a detailed analytical study of the non-Gaussianity in the primordial comoving curvature perturbation, 
as well as its effects on power asymmetry for a single field inflation in the framework of loop quantum cosmology, in which a quantum 
bouncing era prior to inflation exists. The quantum effects naturally generate an excited state on the primordial comoving curvature perturbation 
rather than the usual BD vacuum state at the onset of slow-roll inflation. We have also showed that the excited state may produce enhanced 
effects at large scales, but reduces to the BD state at small scales. With this excited state, we calculate explicitly underlying non-Gaussinity of primordial 
curvature perturbations. It has been shown that the amplitude of the non-Gaussianity in a squeezed limit due to these effects is small at observable scales. 
This is due to the fact that the squeezed limit involves two very different scales. While one of the scales is large, the other is definitely small 
with suppressed effects. We have further shown that the non-Gaussianity can still be enhanced if we consider a superhorizon mode that 
couples to the observable modes at large scales. It is this enhanced non-Gaussianity that leads to a modulation on the isotropic primordial 
power spectrum at observed scales, and thus an explanation of the power asymmetry observed in the CMB spectrum can be naturally concluded.

\section*{Acknowledgements}
We would like to thank Wen Zhao for valuable comments and suggestions. This work is supported in part by Chinese NSF Grants, Nos. 11375153 (A.W.), 11675145(A.W.), 11675143 (T.Z.), and 11105120 (T.Z.). K.K. was supported by the Baylor University Summer Sabbatical Research Program.

\end{document}